\documentclass[12pt]{article}
\usepackage{amsfonts,euscript}
\begin{document}

\title{\bf The $\theta$-twistor versus the supertwistor}
\author {A.A.~Zheltukhin
\\
{\normalsize 
Kharkov Institute of Physics and Technology, 61108 Kharkov, Ukraine,}\\
{\normalsize 
 Fysikum, AlbaNova, University of Stockholm, SE-10691 Stockholm, Sweden}
}                                            
\date{}

\maketitle

\begin{abstract}
We introduce the $\theta$-twistor which is a new supersymmetric generalization of the Penrose twistor and is also alternative to the supertwistor. The $\theta$-twistor is a triple of {\it spinors} including the spinor $\theta$ extending the Penrose's double of spinors. 
Using the $\theta$-twistors yields an infinite chain of  massless higher spin chiral supermultiplets $(\frac{1}{2},1),\, (1, \frac{3}{2}),\,(\frac{3}{2},2),...,(S, S+\frac{1}{2})$ generalizing the known scalar $(0,\frac{1}{2})$ supermultiplet. 
\end{abstract}

\section{Introduction}

The (super)twistor tools  \cite{PR}, \cite{Fbr}, \cite{Witt1} essentialy simplify
the calculation of multigluon amplitudes 
 \cite{Witt2}, \cite{Nai}, \cite{RSV}, \cite{GKh} and sharpen a traditional interest to the search of the twistor role in physics and mathematics
\cite{ADHM}, \cite{CFGT}, 
\cite{Shir}, 
\cite{BC}, \cite{Ber}, \cite{Sieg1}, \cite{Sieg2}.
The supertwistor \cite{Fbr} is a triple including two commuting spinors and the anticommuting scalar $\eta=\nu^\alpha\theta_\alpha$ fixing the contribution of the spinor superspace coordinate $\theta_\alpha$ by only its projection on the Penrose spinor $\nu^\alpha$ \cite{PR}. 
As a result, the chiral supermultiplet in the supertwistor description looses its auxiliary $F$-field
and supersymmetry transformations close on the mass-shell of the Dirac field.
This problem is solved here using  an alternative  way for the twistor supersymmetrization that preserves  all the $\theta_\alpha$ components by the introducing a new super triple in  $D=4\, N=1$ superspace named the $\theta$-twistor who includes $three$ spinors forming a nonlinear supersymmetry representation  \cite{Z}.
We establish that both the $\theta$-twistor and supertwistor appear as the general solutions of two different supersymmetric and Lorentz covariant constraints generalizing the known chirality constraint 
in the superspace extended by  the  Penrose's spinor $\nu_\alpha$. Using the $\theta$-twistor restores the desired $F$-field at the chiral supermultiplet $(0,\frac{1}{2})$. Moreover, it makes possible to reveal an infinite chain of massless higher spin chiral supermultiplets $(\frac{1}{2},1),\, (1, \frac{3}{2}),\,(\frac{3}{2},2),\, ......., (S, S+\frac{1}{2})$ generalizing the well-known scalar supermultiplet $(0,\frac{1}{2})$. 

\section{The supertwistor}

A commuting Weyl spinor  $\nu_{\alpha}$ belonging to the Penrose's  spinor double 
$(\nu_{\alpha}, {\bar\omega}^{\dot\alpha})$  is inert  under the transformations  of  $D=4 \, N=1$ supersymmetry 
\begin{equation}\label{1}
\begin{array}{c}
\delta\theta_\alpha=\varepsilon_\alpha,\quad 
\delta x_{\alpha\dot\alpha}=
2i(\varepsilon_{\alpha}\bar\theta_{\dot\alpha}-
\theta_{\alpha}\bar\varepsilon_{\dot\alpha}), \quad \delta\nu_{\alpha}=0.
\end{array}
\end{equation}
The supertwistor \cite{Fbr} is naturally introduced  starting  from  the complex superspace 
$(y_{\alpha\dot\alpha},\theta_{\alpha}, \bar\theta_{\dot\alpha})$
 and  using the supersymmetric  Cartan-Volkov differential form  $\omega_{\alpha\dot\alpha}$
\begin{equation}\label{2} 
\omega_{\alpha\dot\alpha}=dy_{\alpha\dot\alpha}+ 4id\theta_{\alpha}\bar\theta_{\dot\alpha},\quad
 y_{\alpha\dot\alpha}\equiv x_{\alpha\dot\alpha}-2i\theta_{\alpha}\bar\theta_{\dot\alpha}. 
\end{equation}
If we have the invariant vector differential one-form $\omega_{\alpha\dot\alpha}$  (\ref{2}) and Weyl spinors $\nu_{\alpha},\bar\nu_{\dot\alpha}$ one can construct the scalar invariant form $s=(\nu\omega\bar\nu)$ \cite{UZ},\cite{Z2} that may be presented as  
\begin{equation}\label{3}
s\equiv(\nu\omega\bar\nu)=s(Z,d\bar Z)=-iZ_{\cal A}d\bar Z^{\cal A},
\end{equation}
 where the triples $Z_{\cal A}$ and $\bar Z^{\cal A}$ unify $\nu^{\alpha},\bar\nu_{\dot\alpha}$ with the composite coordinates $q_{\alpha}, \bar q_{\dot\alpha}, \eta, \bar\eta$  
\begin{equation}\label{4}
\begin{array}{c}
Z_{\cal A}\equiv(-iq_{\alpha},\bar\nu^{\dot\alpha}, 2\bar\eta),
\quad
\bar Z^{\cal A}\equiv(\nu^{\alpha},i\bar q_{\dot\alpha}, 2\eta),
\\[0.2cm] 
\eta \equiv \nu^{\alpha}\theta_{\alpha},\quad 
\bar q_{\dot\alpha}= (q_{\alpha})^*\equiv \nu^{\alpha}y_{\alpha\dot\alpha}
=\nu^{\alpha}x_{\alpha\dot\alpha}- 2i\eta\bar\theta_{\dot\alpha}.
\end{array}
\end{equation}
 The triples $Z_{\cal A}$ and $\bar Z^{\cal A}$ coincide with the supertwistor and its c.c. first proposed in \cite{Fbr} to be  a supersymmetric generalization of the projective Penrose twistor. Then the supertwistor space is defined as a complex projective superspace equipped with the invariant bilinear form $s(Z, \bar Z')$ 
\begin{equation}\label{5}
\begin{array}{c}
s(Z, \bar Z')\equiv -iZ_{\cal A}\bar Z'^{\cal A}= -q_{\alpha}\nu'^{\alpha}+ \bar\nu^{\dot\alpha}\bar q'_{\dot\alpha} -4i\bar\eta\eta'=0,
\end{array}
\end{equation}
 where the  triple   $\bar Z'^{\cal A}$ is given by  (\ref{4}) with $\nu'$
substituted for $\nu$
\begin{equation}\label{6}
\begin{array}{c}
\bar Z'^{\cal A}\equiv (\nu'^{\alpha},i\bar q'_{\dot\alpha}, 2\eta'),
\quad 
\bar q'_{\dot\alpha}=\nu'^{\alpha}y_{\alpha\dot\alpha}, \quad \eta'=
\nu'^{\alpha}\theta_{\alpha}.
\end{array}
\end{equation}
It was shown in \cite{Fbr} that the null form (\ref{5}) is invariant under the superconformal symmetry. 
Below we propose an alternative supersymmetric generalization of the Penrose twistor.

\section{The $\theta$-twistor}

Here we introduce an alternative supersymmetric triple including $three$ spinors. This  possibility is
provided by the existence of 
the  new composite spinor $l_{\alpha}$ produced by the right multiplication of $y_{\alpha\dot\alpha}$  (\ref{2}) by $\bar\nu^{\dot\alpha}$ contrarily  to the left multiplication generating  $\bar q_{\dot\alpha}$ (\ref{4})
\begin{equation}\label{7}
l_{\alpha}\equiv y_{\alpha\dot\alpha}\bar\nu^{\dot\alpha}
=x_{\alpha\dot\alpha}\bar\nu^{\dot\alpha}- 2i\theta_{\alpha}\bar\eta, \quad 
l_{\alpha}=q_{\alpha}- 4i\theta_{\alpha}\bar\eta.
\end{equation}
The transformation law of $l_{\alpha}$ (\ref{7}) under the supersymmetry (\ref{1}) is nonlinear
\begin{equation}\label{8}
{\delta l}_{\alpha}=-4i\theta_{\alpha}(\bar\nu^{\dot\beta}\bar\varepsilon_{\dot\beta}), \quad
\delta\theta_{\alpha}=\varepsilon_{\alpha}, \quad
\delta\bar\nu_{\dot\alpha}=0
\end{equation}
and yields a new supersymmetry representation formed by a complex spinor triple $\Xi_{\cal A}$
\begin{equation}\label{9}
\Xi_{\cal A}\equiv(-il_{\alpha},\bar\nu^{\dot\alpha},\theta^{\alpha}),\quad
\bar\Xi^{\cal A}\equiv(\Xi_{\cal A})^*=(\nu^{\alpha},i{\bar l}_{\dot\alpha},\bar\theta^{\dot\alpha}).
\end{equation}
which we call the $\theta$-twistor. 
 The square form (\ref{5}) expressed in terms of $\Xi_{\cal A}$ and $\bar\Xi^{\cal A}$ (\ref{9})
\begin{equation}\label{10}
\begin{array}{c}
s(Z,\bar Z)\equiv
-iZ_{\cal A}\bar Z'^{\cal A}=\tilde s(\Xi,\bar\Xi'), 
\\[0.2cm]
\tilde s(\Xi,\bar\Xi')\equiv
-i\Xi_{\cal A}\circ\bar\Xi'^{\cal A}
=  -l_{\alpha}\nu'^{\alpha}+ 
\bar\nu^{\dot\alpha}\bar l'_{\dot\alpha} - ig_{\alpha\dot\alpha}
\theta^{\alpha}\bar\theta^{\dot\alpha}=0, \quad 
g_{\alpha\dot\alpha}\equiv 
4\nu'_{\alpha}\bar\nu_{\dot\alpha}
\end{array}
\end{equation} 
 becomes a nonlinear form in  the $\theta$-twistor space invariant under supersymmetry and Lorentz symmetry. The generators of the supersymmetry transformations (\ref{8}) take the form \begin{equation}\label{48/1}
\begin{array}{c}
Q^{\alpha}=\frac{\partial}{\partial\theta_{\alpha}}+4i\nu^{\alpha}(\bar\theta_{\dot\beta}\frac{\partial}{\partial\bar l_{\dot\beta}}),
 \quad
 \bar Q^{\dot\alpha}\equiv -(Q^{\alpha})^*=\frac{\partial}{\partial\bar\theta_{\dot\alpha}}+ 4i\bar\nu^{\dot\alpha}(\theta_{\beta}\frac{\partial}{\partial l_{\beta}})
\end{array}
\end{equation}
with their anticommutator closed  by the vector  generator $ P^{\dot\beta\alpha}=(\bar\nu^{\dot\beta}\frac{\partial}{\partial l_{\alpha}}+
\nu^{\alpha}\frac{\partial}{\partial\bar l_{\dot\beta}})$ 
\begin{equation}\label{49/1}
\begin{array}{c}
\{ Q^{\alpha}, \bar Q^{\dot\beta}\}= 4iP^{\dot\beta\alpha},\, \quad
[ Q^{\gamma},P^{\dot\beta\alpha}]=[\bar Q^{\dot\gamma},P^{\dot\beta\alpha}]= 
\{ Q^{\gamma},Q^{\beta}\}= \{\bar Q^{\dot\gamma}, \bar Q^{\dot\beta}\}=0.    
\end{array}
\end{equation}
 It is  easy to see that  the nonlinear null form (\ref{10}) defining  the $\Xi$-space is also invariant under the scaling and phase transformations of the $\theta$-twistor components 
\begin{equation}\label{52/2}
\begin{array}{c}
l'_{\beta}=e^{\varphi}l_{\beta}, 
\quad
{\bar l}'_{\dot\beta}=e^{\varphi*}{\bar l}_{\dot\beta},
 \quad
\nu'_{\beta}=e^{-\varphi}\nu_{\beta} ,
\quad 
\bar\nu'_{\dot\beta}=e^{-\varphi*}\bar\nu_{\dot\beta},
\\[0.2cm]
\theta'_{\beta}=e^{\varphi}\theta_{\beta}, 
\quad
\bar\theta'_{\dot\beta}=e^{\varphi*}\bar\theta_{\dot\beta},
\end{array}
\end{equation}
 described  by the complex parameter $\varphi=\varphi_{R}+i\varphi_{I}$.  Also 
(\ref{10}) is  invariant under the independent $\gamma_5$ rotations of $\theta_{\beta}$ and ${\bar\theta}_{\dot\beta} $  
\begin{equation}\label{50}
\theta'_{\beta}=e^{i\lambda}\theta_{\beta}, \quad 
\bar\theta'_{\dot\beta}=e^{-i\lambda}\bar\theta_{\dot\beta}. 
\end{equation}
There is an alternative way to reveal the supertwistors and the $\theta$-twistors. One can observe that supertwistor appears as the general solution of the supersymmetric constraints 
\begin{equation}\label{61/1}
\begin{array}{c}
{\bar D}^{\dot\alpha}F(x,\theta,\bar\theta)=0 \longrightarrow \, 
F=F(y,\theta),
\\[0.2cm]
\nu_{\alpha}D^{\alpha}F(y,\theta,\nu)=0  \longrightarrow \, F=F(\bar Z^{\cal A})
\end{array}
\end{equation}
 in the chiral  superspace $ (y_{\alpha\dot\alpha},\,\theta_{\alpha})$ complemented by the even spinor  $\nu_{\alpha}$,
where  $F(\bar Z^{\cal A})$ is superfield depending on the supertwistor  $\bar Z^{\cal A}$.
 Conversely,  the $\theta$-twistor is associated with the general solution of other supersymmetric constraints in the chiral space complemented by $\bar\nu_{\dot\alpha}$ 
\begin{equation}\label{64/1}
\begin{array}{c}
{\bar D}^{\dot\alpha}F(x,\theta,\bar\theta)=0 \longrightarrow \, F=F(y,\theta),
\\[0.2cm]
\bar\nu_{\dot\alpha}\frac{\partial}{\partial x_{\alpha\dot\alpha}}F(y,\theta, \bar\nu)=0  
\longrightarrow \, F=F(\Xi_{\cal A}).
\end{array}
\end{equation}
The object of our further investigation is the space of superfunctions  $F(\Xi_{\cal A})$ ({\ref{64/1}) depending on the $\Xi$-triple.

\section{Massless chiral supermultiplets of higher spin fields}

The superfields $F(\bar Z^{\cal A})$ and $F(\Xi_{\cal A})$ describe  massless supermultiplets because they satisfy to  the Klein-Gordon equations 
\begin{equation}\label{67/1}
\partial_{m}\partial^{m}  F(\bar Z)=0, \quad \partial_{m}\partial^{m} F(\Xi)=0, 
 \end{equation}
where $\partial_{m}\equiv(\sigma_{m})_{\dot\alpha\alpha}\partial^{\dot\alpha\alpha}
\equiv(\sigma_{m})_{\alpha\dot\alpha}\frac{\partial}{\partial x_{\alpha\dot\alpha}},\quad 
\partial^{\dot\alpha\alpha}=-\frac{1}{2}\tilde\sigma_{m}^{\dot\alpha\alpha}\partial^m$. 
The component expansion of the analytical function $F(\Xi)$ in the $\Xi$-triple space defined  by 
\begin{equation}\label{73/1}
 F(\Xi)\equiv F(-il_{\alpha},\bar\nu^{\dot\alpha}, \theta^{\alpha})=
f_{0}(-iy_{\beta\dot\beta}\bar\nu^{\dot\beta},  \bar\nu^{\dot\beta})-
 2\theta_{\lambda}f^{\lambda}(-iy_{\beta\dot\beta}\bar\nu^{\dot\beta}, \bar\nu^{\dot\beta}) +\theta^{2}f_{2}(-iy_{\beta\dot\beta}\bar\nu^{\dot\beta}, \bar\nu^{\dot\beta}),
\end{equation}
 where $  \theta_{\alpha}\theta_{\beta}=\frac{1}{2}\varepsilon_{\alpha\beta}\theta^{2}$, preserves the desired auxiliary field $f_{2}$ vanishing in the supertwistor description.  
The contour integral generalizing  the Penrose's supertwistor integral is given by 
\begin{equation}\label{74/1}
\Phi^{\dot\alpha_{1}...\dot\alpha_{2S}}(y,\theta)=\oint(d\bar\nu^{\dot\gamma}\bar\nu_{\dot\gamma})\bar\nu^{\dot\alpha_{1}}...\bar\nu^{\dot\alpha_{2S}}
 F(\bar\nu^{\dot\beta},-i\bar\nu^{\dot\gamma}y_{\beta\dot\gamma}, \theta_{\beta}),
\end{equation}
 where $F(\Xi)$ is supposed to have the  degree of homogeneity equal $-2(S+1)$ and the $\bar\nu$-contour encloses the singularities of $F$ for each fixed point $(y,\theta)$. Inserting  (\ref{73/1}) into (\ref{74/1}) we get
\begin{equation}\label{75/1}
\begin{array}{c}
\Phi^{\dot\alpha_{1}...\dot\alpha_{2S}}(y,\theta)=f_{0}^{\dot\alpha_{1}...\dot\alpha_{2S}}(y)-2\theta_{\lambda}f^{\lambda\dot\alpha_{1}...\dot\alpha_{2S}}(y) +
\theta^{2}f_{2}^{\dot\alpha_{1}...\dot\alpha_{2S}}(y)
\end{array}
\end{equation}
where
\begin{equation}\label{76/1}
\begin{array}{c}
f_{0}^{\dot\alpha_{1}...\dot\alpha_{2S}}(y)=\oint(d\bar\nu^{\dot\gamma}\bar\nu_{\dot\gamma})\bar\nu^{\dot\alpha_{1}}...\bar\nu^{\dot\alpha_{2S}}f_{0}(-iy_{\beta\dot\beta}\bar\nu^{\dot\beta},  \bar\nu^{\dot\beta}),
\\[0.2cm]
f^{\lambda\dot\alpha_{1}...\dot\alpha_{2S}}(y)= \oint(d\bar\nu^{\dot\gamma}\bar\nu_{\dot\gamma})\bar\nu^{\dot\alpha_{1}}...\bar\nu^{\dot\alpha_{2S}}f^{\lambda}(-iy_{\beta\dot\beta}\bar\nu^{\dot\beta}, \bar\nu^{\dot\beta}),
\\[0.2cm]
f_{2}^{\dot\alpha_{1}...\dot\alpha_{2S}}(y)=\oint(d\bar\nu^{\dot\gamma}\bar\nu_{\dot\gamma})\bar\nu^{\dot\alpha_{1}}...\bar\nu^{\dot\alpha_{2S}}f_{2}(-iy_{\beta\dot\beta}\bar\nu^{\dot\beta},  \bar\nu^{\dot\beta})
\end{array}
\end{equation}
with $f^{\lambda\dot\alpha_{1}...\dot\alpha_{2S}}(y)$ and  $f_{0,2}^{\dot\alpha_{1}...\dot\alpha_{2S}}(y)$ satisfying the chiral Dirac equations
\begin{equation}\label{77/1}
\partial_{\alpha\dot\alpha_{k}}
f^{\lambda\dot\alpha_{1}..\dot\alpha_{k}..\dot\alpha_{2S}}(x)= \partial_{\alpha\dot\alpha_{k}}
f_{0,2}^{\dot\alpha_{1}..\dot\alpha_{k}..\dot\alpha_{2S}}(x)=0, 
\quad (k=1,2,...,2s).
\end{equation}
The further expansion of $\Phi^{\dot\alpha_{1}...\dot\alpha_{2S}}(y,\theta)$ 
(\ref{75/1})  at the point $x_{m}$ is  given by 
\begin{equation}\label{80/1}
\begin{array}{c}
\Phi^{\dot\alpha_{1}...\dot\alpha_{2S}}(y,\theta)=f_{0}^{\dot\alpha_{1}...\dot\alpha_{2S}}(x)
-2\theta_{\lambda}f^{\lambda\dot\alpha_{1}...\dot\alpha_{2S}}(x) -
2i\theta_{\gamma}\bar\theta_{\dot\gamma}\partial^{\dot\gamma\gamma} f_{0}^{\dot\alpha_{1}...\dot\alpha_{2S}}(x)
\\[0.2cm]
-2i\theta^{2}\bar\theta_{\dot\gamma}\partial^{\dot\gamma\lambda}f_{\lambda}^{\dot\alpha_{1}...\dot\alpha_{2S}}(x)+\theta^{2}f_{2}^{\dot\alpha_{1}...\dot\alpha_{2S}}(x),
\end{array}
\end{equation}
where the term $\frac{1}{2}\theta^{2}{\bar\theta}^{2}\partial^{\dot\gamma\gamma}\partial_{\gamma\dot\gamma}f_{0}^{\dot\alpha_{1}...\dot\alpha_{2S}}(x)$ was droped because of the zero mass constraint 
(\ref{67/1})
\begin{equation}\label{81/1}
\begin{array}{c}
\Box\Phi^{\dot\alpha_{1}...\dot\alpha_{2S}}(y,\theta)=0 \longrightarrow \Box f_{0}^{\dot\alpha_{1}...\dot\alpha_{2S}}(x)=0.
\end{array}
\end{equation}
For sewing together of these results with the well-known case $S=0$ corresponding to the scalar supermultiplet let us rename the component $f$-fields by the letters used in \cite{WB} 
\begin{equation}\label{82/1}
\begin{array}{c}
f_{0}^{\dot\alpha_{1}...\dot\alpha_{2S}}={\sqrt{2}}A^{\dot\alpha_{1}...\dot\alpha_{2S}}
\equiv{\sqrt{2}}A^{...},
\quad
f_{2}^{\dot\alpha_{1}...\dot\alpha_{2S}}=\sqrt{2}F^{\dot\alpha_{1}...\dot\alpha_{2S}}
\equiv\sqrt{2}F^{...},
\\[0.2cm]
f_{\lambda}^{\dot\alpha_{1}...\dot\alpha_{2S}}=\psi_{\lambda}^{\dot\alpha_{1}...\dot\alpha_{2S}} 
\equiv\psi_{\lambda}^{...},
\end{array}
\end{equation}
where $(...)\equiv (\dot\alpha_{1}...\dot\alpha_{2S})$. 
Then we find the superfield $\frac{1}{\sqrt{2}}\Phi^{...}(y,\theta)$ to describe the massless chiral multiplet \cite{WB} for the case $S=0$. 
For  $S\neq0$ the superfield (\ref{75/1}) represents  the chiral supermultiplets of  massless higher spin fields with the particle spin content $$(\frac{1}{2},1),\, (1, \frac{3}{2}),\,(\frac{3}{2},2),\, ......., (S, S+\frac{1}{2})$$  accompanied by the  corresponding  auxiliary fields for any integer or half-integer spin  $ S=\frac{1}{2},1, \frac{3}{2},2, ....$.   
The supersymmetry transformations for the higher spin multiplet (\ref{75/1}) presented  in the notations (\ref{82/1}) take  the form 
\begin{equation}\label{84/1}
\begin{array}{c}
\delta A^{...}=\sqrt{2}\varepsilon^{\lambda}\psi_{\lambda}^{...}, \quad
\delta F^{...}=i\sqrt{2}(\bar\varepsilon{\tilde\sigma}_{m}
\partial^{m}\psi^{...})
\\[0.2cm]
\delta\psi_{\lambda}^{...}=i\sqrt{2}(\sigma_{m}\bar\varepsilon)_{\lambda}\partial^{m}A^{...}+ 
\sqrt{2}\varepsilon_{\lambda}F^{...}
\end{array}
\end{equation}
 and coincide  with the transformation rules for the $S=0$ chiral multiplet of the weight 
$n=\frac{1}{2}$  \cite{WB}  if we put   $A^{...}=A,\, F^{...}=F$ and $\psi_{\lambda}^{...}= \psi_{\lambda}$  in the  relations (\ref{84/1}).

As it was noted the $\theta$-twistor superspace is invariant under the axial rotations (\ref{50}) and one can consider these phase transformations as inducing the $R$-symmetry transformations for the superfield  $F(\Xi)$ (\ref{73/1})
\begin{equation}\label{85/1}
\begin{array}{c}
 F'(-il_{\alpha}, \bar\nu^{\dot\alpha},  e^{i\varphi}\theta^{\alpha})=e^{2in\varphi}F(-il_{\alpha}, \bar\nu^{\dot\alpha}, \theta^{\alpha}),
\end{array}
\end{equation}
where $n$ is the correspondent $R$ number. Then  taking into account the representation (\ref{74/1}) we get 
  the $R$-symmetry transformation of the generalized chiral superfield 
$\Phi^{\dot\alpha_{1}...\dot\alpha_{2S}}(y,\theta)$
\begin{equation}\label{86/1}
\Phi'^{\dot\alpha_{1}...\dot\alpha_{2S}}(y,\theta)=
e^{2in\varphi}\Phi^{\dot\alpha_{1}...\dot\alpha_{2S}}(y,e^{-i\varphi}\theta).
\end{equation}
So, one can expect that new renormalizable Lagrangians may be  constructed  using these higher spin superfields.   

\section{Conclusion}

Proposed is the new supersymmetric generalization of the Penrose twistor generating the $\theta$-twistor. As a new mathematical object the  $\theta$-twistor deserves to be studied both  in its  own rights and in further physical applications. A creative charge of the $\theta$-twistor was here illustrated by the production of an infinite chain of higher spin chiral supermultiplets generalizing the massless scalar supermultiplet.
  The new  chiral superfields  may be used for the construction of new physically interesting models because the studied here  $D=4 \, N=1$ example has the direct generalization both to the case of extended supersymmetries  by the change 
$(\theta_{\alpha},\,{\bar\theta}_{\dot\alpha})\rightarrow \theta^{i}_{a}$ and to higher dimensions $D=2,3,4(mod8)$ by analogy with the supertwistors \cite{Fbr}, \cite{Witt1}, \cite{BZ}.

\section{Acknowledgements}

The author is grateful to  Fysikum at the Stockholm University for kind hospitality and I. Bengtsson, F. Hassan, E. Langmann and  U. Lindstr{\"o}m  for useful discussions.
I thank  J. Lukierski for the reference  \cite{FIL}, where other chiral HS supermultiplets were proposed.
This work was partially supported by the grant of the Royal Swedish Academy of Sciences.

\end{document}